\documentclass[conference]{IEEEtran}
\usepackage[hang,flushmargin]{footmisc}
\usepackage{times}

\usepackage{setspace}
\usepackage{multirow}
\usepackage{hhline}

\usepackage{wrapfig}
\usepackage[bookmarks=false]{hyperref}
\usepackage{amssymb}
\usepackage{amsthm}
\let\labelindent\relax
\usepackage{enumitem}
\usepackage{listings}
\usepackage{epstopdf}
 \usepackage[noadjust]{cite}
\usepackage{amssymb}

\usepackage{xspace}

\usepackage{csquotes}
\usepackage[boxed,ruled,vlined,linesnumbered]{algorithm2e}
\usepackage{url}
\usepackage{framed}
\usepackage{caption}

\usepackage{fancyhdr}
\usepackage{subcaption}

\usepackage[none]{hyphenat}
\usepackage{times}
\usepackage{graphicx}
\newcommand{\B}{\vspace*{-\smallskipamount}}
\newcommand{\BB}{\vspace*{-\medskipamount}}
\newcommand{\BBB}{\vspace*{-\bigskipamount}}

\SetAlgoSkip{}

\IEEEoverridecommandlockouts

\usepackage{times,amsmath,epsfig,amsthm}

\usepackage{balance}
\usepackage{booktabs}
\usepackage{setspace}
\usepackage{enumitem}
\usepackage{multirow}
\usepackage{hhline}
\usepackage{wrapfig}
\usepackage{amssymb}
\usepackage{xspace}
\usepackage{mdframed}
\usepackage{csquotes}
\usepackage[boxed,ruled,vlined,linesnumbered]{algorithm2e}
\usepackage{url}
\usepackage{framed}
\usepackage[utf8]{inputenc}
\usepackage[english]{babel}

\usepackage{hyperref}
\usepackage{cite}
\usepackage{caption}
\usepackage{subcaption}
\usepackage{times}
\usepackage{graphicx}
\usepackage{times,amsmath,epsfig}
\usepackage{amsthm}
\usepackage{soul}

\SetAlgoSkip{}
\SetAlFnt{\small}
\SetAlCapFnt{\small}
\SetAlCapNameFnt{\small}
\SetAlCapHSkip{0pt}
\IncMargin{-\parindent}

\usepackage{tikz}
\newcommand\encircle[1]{%
\tikz[baseline=(X.base)]
   \node (X) [draw, shape=circle, inner sep=-1.5pt, fill=black, text=white] {\strut #1};}

\begin{document}


\title{\textsc{IoT Notary}: Sensor Data Attestation in Smart Environment}
\author{\IEEEauthorblockN{Nisha Panwar, Shantanu Sharma, Guoxi Wang, Sharad Mehrotra, Nalini Venkatasubramanian, \\ Mamadou H. Diallo, and Ardalan Amiri Sani \thanks{
\textbf{Accepted in IEEE International Symposium on Network Computing and Applications (NCA), 2019.} For the final version, please refer to the conference proceeding.  \newline This work is based on research sponsored by DARPA under agreement number FA8750-16-2-0021 and partially supported by NSF grants 1527536 and 1545071. The U.S. Government is authorized to reproduce and distribute reprints for Governmental purposes notwithstanding any copyright notation thereon. The views and conclusions contained herein are those of the authors and should not be interpreted as necessarily representing the official policies or endorsements, either expressed or implied, of DARPA or the U.S. Government.} }
\IEEEauthorblockA{University of California, Irvine, California, USA.\BB}}


\maketitle

\begin{abstract}
Contemporary IoT environments, such as smart buildings, require end-users to trust data-capturing rules published by the systems. There are several reasons why such a trust is misplaced --- IoT systems may violate the rules deliberately or IoT devices may transfer user data to a malicious third-party due to cyberattacks, leading to the loss of individuals' privacy or service integrity. To address such concerns, we propose \textsc{IoT Notary}, a framework to ensure trust in IoT systems and applications. \textsc{IoT Notary} provides secure log sealing on live sensor data to produce a verifiable `proof-of-integrity,' based on which a verifier can attest that captured sensor data adheres to the published data-capturing rules. \textsc{IoT Notary} is an integral part of TIPPERS, a smart space system that has been deployed at UCI to provide various real-time location-based services in the campus. \textsc{IoT Notary} imposes nominal overheads for verification, thereby users can verify their data of one day in less than two seconds.
\end{abstract}
%
\section{Introduction}
\label{sec:introduction}
While fine-grained continuous monitoring by IoT devices (\textit{e}.\textit{g}., camera and WiFi access-points) offers numerous benefits and empowers existing systems with new capabilities, it also raises several privacy and security concerns (\textit{e}.\textit{g}., smoking habits, gender, and religious belief). To highlight the privacy concern, we first share our experience in building location-based services at UC Irvine using WiFi connectivity data.


\medskip
\noindent\textbf{Use-case: University WiFi data collection.} In our on-going project, entitled TIPPERS~\cite{DBLP:conf/percom/MehrotraKVR16}, we have developed a variety of location-based services based on WiFi connectivity dataset. At UC Irvine, more than 2000 WiFi access-points and four WLAN controllers (managed by the university IT department) provide campus-wide wireless network coverage. Whenever a device connects to the campus WiFi network (through an access-point), the access-point generates Simple Network Management Protocol (SNMP) trap for this association event. Each association event contains access-point-id, $s_i$, user device MAC address, $d_j$, and the time of the association, $t_k$. All SNMP traps $\langle s_i,d_j,t_k\rangle$ are sent to access-point's controllers in realtime. The access-point controller anonymizes device MAC addresses (to preserve the privacy of  users in the campus).

TIPPERS collects WiFi connectivity data from one of the controllers that manage 490 access-points and receives $\langle s_i,d_j,t_k\rangle$ tuples for each connectivity event. However, before receiving any WiFi data, TIPPERS notifies all WiFi users about the data-capture rules by sending emails over a university mailing list. Subsequently, based on WiFi connectivity data $\langle s_i,d_j,t_k\rangle$, TIPPERS provides various realtime applications. Some of these services, \textit{e}.\textit{g}., computing occupancy levels of (regions in) buildings in the form of a live heatmap, require only anonymous data. Other services, \textit{e}.\textit{g}., providing location information (within buildings) or contextualized messaging (to provide messages to a user when he/she is in the vicinity of the desired location), require user's original disambiguated data. To date, over one hundred users have registered into TIPPERS to utilize realtime services. A key requirement imposed by the university in sharing data with TIPPERS is that the system supports provable mechanisms to verify that individuals have been notified prior to their data (anonymized or not) being used for service provisioning. Also, an option for users to opt-out of sharing their WiFi connectivity data with TIPPERS must be supported. If users opt-out, the system must prove to the users that indeed their data was not shared with TIPPERS. TIPPERS use immutable log-sealing to help all users to verify that the captured data is consistent with pre-notified data-capture rules.

Our experience in working with various groups in the campus is that (persistent) location data can be deemed quite sensitive by certain individuals with concerns about the spied upon by the administration or by others. Thus, mechanisms for notification of data-capture rules, secure log-sealing, and verification components made a sub-framework, entitled \textsc{IoT Notary}, which has become an integral part of TIPPERS.

Data-capture concerns in IoT environments are similar to that in mobile computing, where mobile applications may have continuous access to resident sensors on mobile devices. In the mobile setting, data-capture rules and permissions 
are used to control data access, \textit{i}.\textit{e}., which applications have access to which data generated at the mobile device (\textit{e}.\textit{g}., location and contact list) for which purpose and in which context. 
However, in IoT settings, the data-capture framework differs from that in the mobile settings, in two important ways:

\begin{enumerate}[nolistsep,noitemsep,leftmargin=0.2in]
  \item Unlike the mobile setting, where applications can seek user's permission at the time of installation, in IoT settings, there are no obvious mechanisms/interfaces to seek users' preferences about the data being captured by sensors of the smart environment. Recent work~\cite{sup} has begun to explore mechanisms using which environments can broadcast their data-capture rules to users and seek their explicit permissions.
  \item Unlike the mobile setting, users cannot control sensors in IoT settings. While in mobile settings, a user can trust the device operating system not to violate the data-capture rules, in IoT settings, trust (in the environment controlling the sensors) may be misplaced. IoT systems may not be honest or may inadvertently capture sensor data, even if data-capture rules are not satisfied.
\end{enumerate}

We focus on the above-mentioned second scenario and determine ways to provide trustworthy sensing in an untrusted IoT environment. Thus, the users can verify their data captured by IoT environment based on pre-notified data-capture rules. Particularly, we deal with three sub-problems, namely \emph{secure notification} to the user about data-capture rules, \emph{secure (sensor data) log-sealing} to retain \emph{immutable} sensor data, as well as, data-capture rules, and \emph{remote attestation} to verify the sensor data against pre-notified data-capture rules by a user, without being heavily involved in the attestation process.

\smallskip
\noindent\textbf{Our contribution and outline of the paper.} We provide:
\begin{itemize}[nolistsep,noitemsep,leftmargin=0.1in]
\item A user-centric framework (\S\ref{sec:high_level_iot}) to ensure trustworthy data collection in untrusted IoT spaces, entitled \textsc{IoT Notary}. 

\item Two models to inform the user about the data-capture rules (\S\ref{subsec:notification phase}): notice-only model and notice-and-ACK model.

\item A secure log-sealing mechanism (\S\ref{sec:Log Sealing}) implemented by secure hardware that cryptographically retains logs, data-capture rules, sensors' state, and contextual information to generate a \textit{proof-of-integrity} in an immutable fashion. 

\item A secure attestation mechanism (\S\ref{subsec:Attestation Phase}), mixed with SIGMA protocol~\cite{luks}, allowing a verifier (a user or a \emph{non-mandatory} auditor) to securely attest the sealed logs as per the data-capture rules. Implementation results of \textsc{IoT Notary} on the university live WiFi data are provided in \S\ref{sec:Experimental Evaluation}. 
\end{itemize}

\smallskip\noindent\textbf{Full version.} Due to space limitations, we could not describe several details about \textsc{IoT Notary}, which are given in the full version~\cite{fullversion}. These include: future temporal password-based notification method, log retrieval at the verifier using SIGMA, details of the verification phase, throughput and communication cost experiments, and security proofs.

\section{Modeling IoT Data Attestation}
\label{sec:Preliminaries}
\subsection{Entities}
\label{subsec:Entities}
\begin{figure}
  \centering
	\includegraphics[scale=0.36]{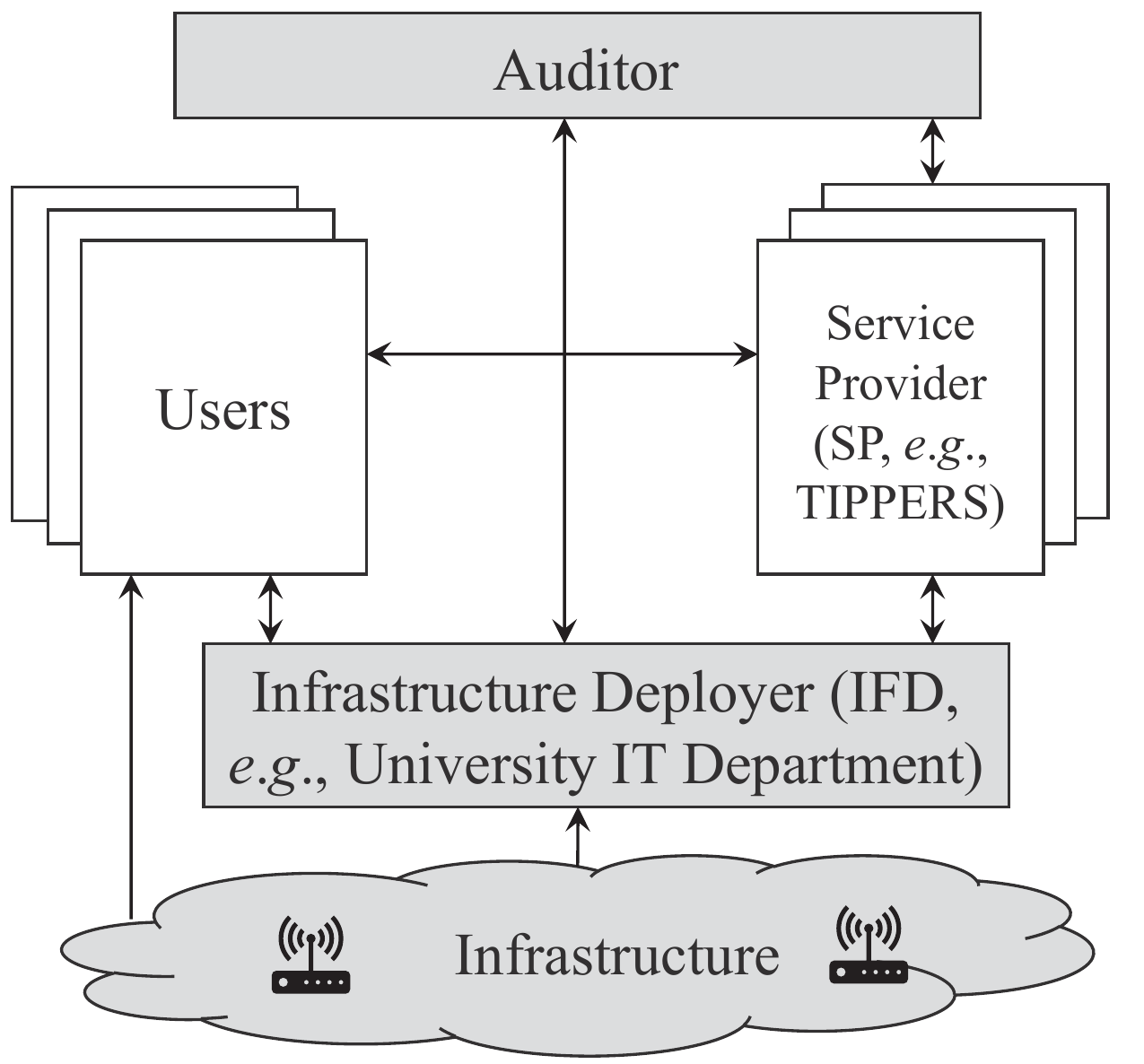}
  \caption{Entities in \textsc{IoT Notary}.}
  \label{fig:entity}
  \BBB
\end{figure}

Our model has the following entities, see Figure~\ref{fig:entity}:

\medskip
\noindent\textbf{Infrastructure Deployer (IFD).} IFD (which is the university IT department in our use-case; see \S\ref{sec:introduction}) deploys and owns a network of $p$ sensors devices (denoted by $s_1, s_2, \ldots, s_p$), which capture information related to users in a space. The sensor devices could be: (\textit{i}) dedicated sensing devices, \textit{e}.\textit{g}., energy meters and occupancy detectors, or (\textit{ii}) facility providing sensing devices, \textit{e}.\textit{g}., WiFi access-points and RFID readers. Our focus is on facility providing sensing devices, especially WiFi access-points that also capture some user-related information in response to services. E.g., WiFi access-points capture the associated user-device-ids (MAC addresses), time of association, some other parameters (such as signal strength, signal-to-noise ratio); denoted by: $\langle d_i, s_j, t_k, \mathit{param} \rangle$, where $d_i$ is the $i^{\mathit{th}}$ user-device-id, $s_j$ is the $j^{\mathit{th}}$ sensor device, $t_k$ is $k^{\mathit{th}}$ time, and $\mathit{param}$ is other parameters (we do not deal with $\mathit{param}$ field and focus on only the first three fields). All sensor data is collected at a controller (server) owned by IFD. The controller may keep sensor data in cleartext or in encrypted form; however, it only sends encrypted sensor data to the service provider.

\medskip
\noindent\textbf{Service Providers (SP).} SP (which is TIPPERS in our use-case; see \S\ref{sec:introduction}) utilizes the sensor data of a given space to provide different \emph{services}, \textit{e}.\textit{g}., monitoring a location and tracking a person. SP receives encrypted sensor data from the controller.

\smallskip
\noindent\underline{\emph{Data-capture rules}}. SP establishes data-capture rules (denoted by a list $\mathcal{\mathit{DC}}$ having different rules $\mathit{dc}_1,\mathit{dc}_2, \ldots, \mathit{dc}_q$). Data-capture rules are conditions on device-ids, time, and space. Each data-capture rule has an associated \emph{validity} that indicates the time during which a rule is valid. Data-capture rules could be to capture user data by default (unless the user has explicitly opted out). Alternatively, default rules may be to opt-out, unless, users opt-in explicitly. Consider a default rule that individuals on the $6^{\mathit{th}}$ floor of the building will be monitored from 9pm to 9am. Such a rule has an associated condition on the time and the id of the sensor used to generate the data. Now, consider a rule corresponding to a user with a device $d_i$ opting-out of data capture based on the previously mentioned rule. Such an opt-out rule would have  conditions on the user-id, as well as, on time and the sensor-id. For sensor data for which a default data-capture rule is opt-in, the captured data is forwarded to SP, if there does not exist any associated opt-out rules, whose associated conditions are satisfied by the sensor data. Likewise, for sensor data where the default is opt-out, the data is forwarded to SP only, if there exists an explicit opt-in condition. We refer to the sensor data to have a \emph{sensor state} ($s_i.\mathit{state}$ denotes the state of the sensor $s_i$) of 1 (or active), if the data can be forwarded to SP; otherwise, 0 (or passive). In the remaining paper, unless explicitly noted, opt-out is considered as the default rule, for simplicity of discussion.

Whenever SP creates a new data-capture rule, SP must send a \emph{notice message} to user devices about the current usage of sensor data (this phase is entitled \emph{notification phase}). SP uses Intel Software Guard eXtension (SGX)~\cite{sgx}, which works as a trusted agent of IFD, for securely storing sensor data corresponding to data-capture rules. SGX keeps all valid data-capture rules in the secure memory and only allows to keep such data that qualifies pre-notified valid data-capture rules; otherwise, it discards other sensor data. Further, SGX creates immutable and verifiable logs of the sensor data (this phase is entitled \emph{log-sealing phase}). The assumption of secure hardware at a machine is rational with the emerging system architectures, \textit{e}.\textit{g}., Intel machines are equipped with SGX~\cite{url1}. However, existing SGX architectures suffer from side-channel attacks, \textit{e}.\textit{g}., cache-line, branch shadow, page-fault attacks~\cite{DBLP:conf/ccs/WangCPZWBTG17}, which are outside the scope of this paper.


\medskip
\noindent\textbf{Users.} Let $d_1,d_2,\ldots,d_m$ be $m$ (user) devices carried by $u_1,u_2,\ldots,u_{m^{\prime}}$ users, where $m^{\prime}\leq m$. Using these devices, users enjoy services provided by SP. We define a term, entitled \emph{user-associated data}. Let $\langle d_i, s_j, t_k\rangle$ be a sensor reading. Let $d_i$ be the $i^{\mathit{th}}$ device-id owned by a user $u_i$. We refer to $\langle d_i, s_j, t_k\rangle$ as user-associated data with the user $u_i$. Users worry about their privacy, since SP may capture user data without informing them, or in violation of their preference (\textit{e}.\textit{g}., when the opt-out was a default rule or when a user opted-out from an opt-in default). Users may also require SP to prove service integrity by storing all sensor data associated with the user (when users have opted-in into services), while minimally being involved in the attestation process and storing records at their sides (this phase is entitled \emph{attestation phase}).

\medskip
\noindent\textbf{Auditor.} An auditor is a \emph{non-mandatory} trusted-third-party that can (periodically) verify entire sensor data against data-capture rules. Note that a user can only verify his/her data, not the entire sensor data or sensor data related to other users, since it may reveal the privacy of other users.

\subsection{Threat Model}
\label{subsec:Threat Model}
We assume that SP and users may behave like adversaries. The adversarial SP may \emph{store} sensor data without informing data-capture rules to the user. The adversarial SP may \emph{tamper} with the sensor data by inserting, deleting, modifying, and truncating sensor readings and secured-logs in the database. By tampering with the sensor data, SP may \emph{simulate} the sealing function over the sensor data to produce secured-logs that are identical to real secured-logs. Thus, the adversary may hinder the attestation process and make it impossible to detect any tampering with the sensor data by the verifier (that may be an auditor or a user). Further, as mentioned before that SP utilizes sensor data to provide services to the user. However, an adversarial SP may provide \emph{false answers} in response to user queries. We assume that the adversarial SP cannot obtain the secret key of the enclave (by any means of side-channel attacks on SGX). Since we assumed that sensors are trusted and cannot be spoofed, we do not need to consider a case when sensors would collude with SP to fabricate the logs.

An adversarial user may \emph{repudiate} the reception of notice messages about data-capture rules. Also, an adversarial user may \emph{impersonate} a real user to retrieve the sensor data and secured-log during the verification phase. Thus, an adversarial user may reveal the privacy of the users by observing sensor data. Also, a user may infer the identity of other users associated with sensor data by potentially launching \emph{frequency-count attacks} (\textit{e}.\textit{g}., by determining which device-ids are prominent).

\subsection{Security Properties}
\label{subsec:Properties}
In the above-mentioned adversarial model, an adversary wishes
to learn the (entire/partial) data about the user, without notifying or by mis-notifying about data-capture rules, such that the user/auditor cannot detect any inconsistency between data-capture rules and stored sensor data at SP. Hence, a secure attestation algorithm must make it detectable, if the adversary stores sensor data in violation of the data-capture rules notified to the user. To achieve a secure attestation algorithm, we need to satisfy the following properties:

\medskip
\noindent\textbf{Authentication.} Authentication is required: (\textit{i}) between SP and users, during notification phase; thus, the user can detect a rogue SP, as well as, SP can detect rogue users, and (\textit{ii}) between SP and the verifier (auditor/user), before sending sensor data to the verifier to prevent any rogue verifier to obtain sensor data. Thus, authentication prevents threats such as impersonation and repudiation. Further, a periodic mutual authentication is required between IFD and SP, thereby discarding rogue sensor data by SP, as well as, preventing any rogue SP to obtain real sensor data.

\medskip
\noindent\textbf{Immutability and non-identical outputs.} We need to maintain immutability of notice messages, sensor data, and the sealing function. Note that if the adversary can alter notice messages after transmission, it can do anything with the sensor data, in which case, sensor data may be completely stored or deleted without respecting notice messages. Further, if the adversary can alter the sealing function, the adversary can generate a proof-of-integrity, as desired, which makes the flawless attestation impossible. The output of the sealing function should not be identical for each sensor reading to prevent an adversary to forge the sealing function (and to prevent the execution of frequency-count attack by the user). Thus, immutability and non-identical outputs properties prevent threats, \textit{e}.\textit{g}., inserting, deleting, modifying, and truncating the sensor data, as well as, simulating the sealing function.

\begin{figure*}[!t]
	\begin{center}
	\includegraphics[scale=0.45]{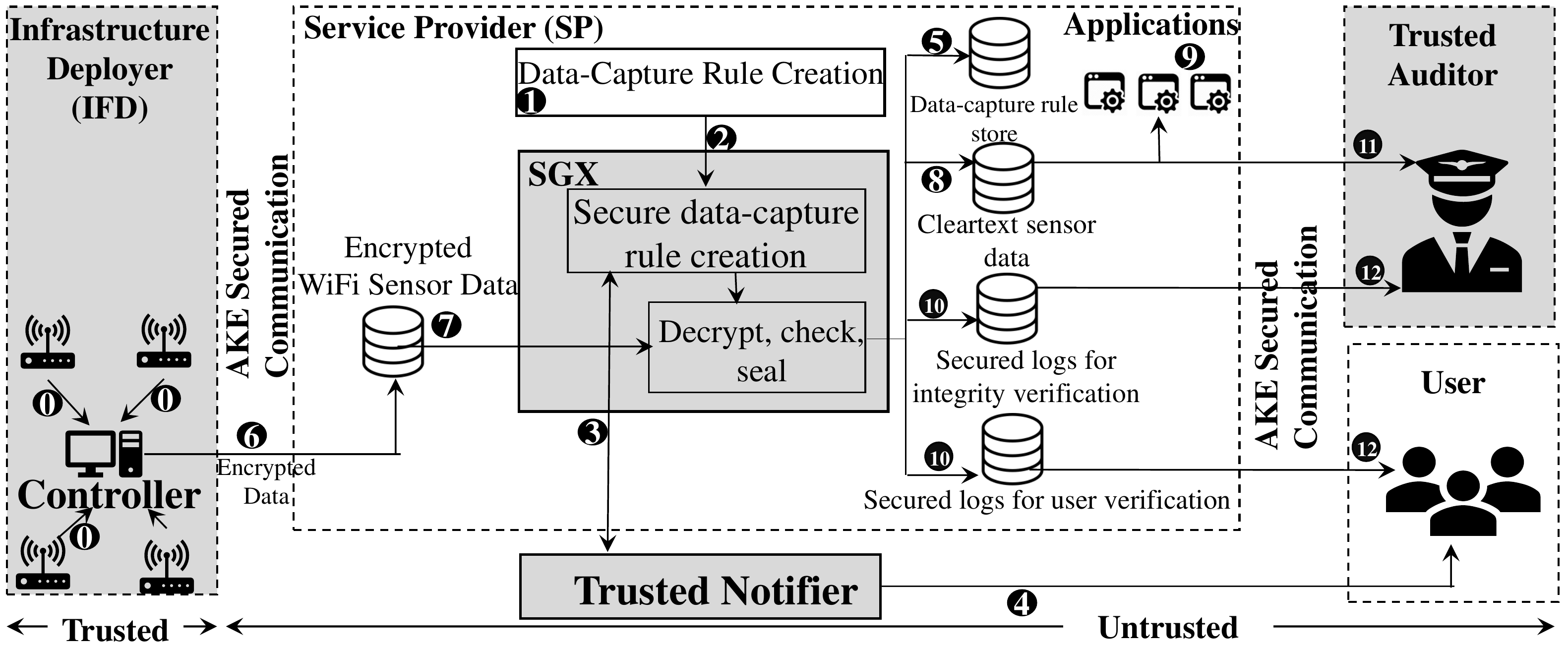}
	\end{center}
    \caption{Dataflow and computation in the protocol. Trusted parts are shown in shaded boxes.}
	\BBB
	\label{fig:Dataflow and computation in the protocol}
\end{figure*}

\medskip
\noindent\textbf{Minimality, non-refutability and privacy-preserving verification.} The verification method must find any misbehavior of SP, during storing sensor data inconsistent with pre-notified data-capture rules. However, if the verifiers wish to verify a subset of the sensor data, then they should not verify the entire sensor data. Thus, SP should send a minimal amount of sensor data to the verifier, enabling them to attest what they wish to attest. Further, the verification method: (\textit{i}) cannot be refuted by SP, and (\textit{ii}) should not reveal any additional information to the user about all the other users during the verification process. These properties prevent SP to store only sensor data that is consistent with the data-capture rules notified to the user. Further, these properties preserve the privacy of other users during attestation and impose minimal work on the verifier.

\subsection{Assumptions}
\label{subsec:The Attestation Problem}
This section presents assumptions, we made, as follows:

\begin{enumerate}[noitemsep,nolistsep,leftmargin=0.01in]
  \item The sensor devices are assumed to be computationally-inefficient to locally generate a verifiable log for the continuous data stream as per the data-capture rules.
  \item Sensor devices are tamper-proof, and they cannot be replicated/spoofed (\textit{i}.\textit{e}., two devices cannot have an identical id). In short, we assume a correct identification of sensors, before accepting any sensor-generated data at the controller at IFD, and it ensures that no rogue sensor device can generate the data on behalf of an authentic sensor. Further, we assume that an adversary cannot deduce any information from the dataflow between a sensor and the controller. Recall that in our setting the university IT department collects the entire sensor data from their owned and deployed sensors, before sending it to TIPPERS.
  \item We assume the existence of an authentication protocol between the controller and SP, so that SP receives sensor data only from authenticated and desired controller.
  \item The communication channels between SP and users, as well as, between SP and auditor are insecure. Thus, our solution incorporates an authenticated key exchange based on SIGMA protocol (which protects sender identity). When the verifier's identity is proved, the cryptographically sealed logs are sent to the verifier.

  \item By any side-channel attacks on SGX, one cannot tamper with SGX and retrieve the secret-key of SGX. (Otherwise, the adversary can simulate the sealing process.)
\end{enumerate}

\section{\textsc{IoT Notary}}
\label{sec:high_level_iot}

This section presents an overview of the three phases and dataflow among different entities and devices, see Figure~\ref{fig:Dataflow and computation in the protocol}.

\medskip
\noindent\textbf{Notification phase: SP to Users messages.} This is the first phase that notifies users about data-capture rules for the IoT space using notice messages (in a verifiable manner for later stages). Such messages can be of two types: (\textit{i}) notice messages, and (\textit{ii}) notice-and-acknowledgment messages. SP establishes (the default) data-capture rules and informs trusted hardware (\encircle{1}). Trusted hardware securely stores data-capture rules (\encircle{2}, \encircle{5}) and informs the \emph{trusted notifier} (\encircle{3}) that transmits the message to all users (\encircle{4}). Only notice messages need a trusted notifier to transmit the message (see \S\ref{subsec:notification phase}).

\medskip
\noindent\textbf{Log-sealing phase: Sensor devices to SP messages.}
Each sensor sends data to the controller (\encircle{0}). The controller receives the correct data, generated by the actual sensor, as per our assumptions (and settings of the university IT department). The controller sends encrypted data to SP (\encircle{6}) that authenticates the controller using any existing authentication protocol, before accepting data. Trusted hardware (Intel SGX) at SP reads the encrypted data in the enclave (\encircle{7}).

\smallskip
\noindent\textit{\underline{Working of the enclave.}} The enclave decrypts the data and checks against the pre-notified data-capture rules. Recall that the decrypted data is of the format: $\langle d_i, s_j, t_k \rangle$, where $d_i$ is $i^{\mathit{th}}$ user-device-id, $s_j$ is the $j^{\mathit{th}}$ sensor device, and $t_k$ is $k^{\mathit{th}}$ time. After checking each sensor reading, the enclave adds a new field, entitled {\em sensor (device) states}. The sensor state of a senor $s_j$ is denoted by $s_j.\mathit{state}$, which can be \texttt{active} or \texttt{passive}, based on capturing user data. For example, $s_j.\mathit{state}$ = \texttt{active} or (\texttt{1}), if data captured by the sensor $s_j$ satisfies the data-capture rules; otherwise, $s_j.\mathit{state}$ = \texttt{passive} or (\texttt{0}). For all the sensors whose $\mathit{state} = 0$, the enclave deletes the data. Then, the enclave cryptographically seals sensor data, regardless of the sensor state, and provides cleartext sensor data of the format: $\langle d_i, s_j, s_j.\mathit{state}=1,t_k \rangle$ to SP (\encircle{8}) that provides services using this data (\encircle{9}). Note that the cryptographically sealed logs and cleartext sensor data are kept at untrusted storage of SP (\encircle{8}, \encircle{10}).

\medskip
\noindent\textbf{Verification phase: SP to verifier messages.} In our model, an auditor and a user can verify the sensor data. The auditor can verify the entire/partial sensor data against data-capture rules by asking SP to provide cleartext sensor data and cryptographically sealed logs (\encircle{8}, \encircle{10}). The users can also verify their own data against pre-notified messages or can verify the results of the services provided by SP using only cryptographically sealed logs (\encircle{12}). Note that using an underlying authentication technique (as per our assumptions), auditor/users and SP authenticate each other before transmitting data from SP to auditor/users.

\section{Attestation Protocol}
This section presents three phases of attestation protocol.

\medskip
\noindent\textbf{Preliminary Setup Phase.} We assume a preliminary setup phase that distributes public keys ($\mathit{PK}$) and private keys ($\mathit{PR}$), as well as, registers user devices into the system. The trusted authority (which is the university IT department in our setup of TIPPERS) generates/renews/revokes keys used by the secure hardware enclave (denoted by $\langle \mathit{PK}_E, \mathit{PR}_E\rangle$) and the notifier (denoted by $\langle \mathit{PK}_N, \mathit{PR}_N\rangle$). The keys are provided to the enclave during the secure hardware registration process. Also, $\langle \mathit{PK}_{\mathit{di}}, \mathit{PR}_{\mathit{di}}\rangle$ denotes keys of the $i^{\mathit{th}}$ user device. \noindent\emph{Usages of keys}: The controller uses $\mathit{PK}_E$ to encrypt sensor readings before sending to SP. $\mathit{PR}_E$ is also used by the enclave to write encrypted sensor logs and decrypt sensor readings. $\mathit{PK}_N$ is used during the notification phase by SGX to send an encrypted message to the notifier. User device's keys are used during device registration, as given below.

We assume a registration process during which a user identifies herself to the underlying system. For instance, in a WiFi network, users are identified by their mobile devices, and the registration process consists of users providing the MAC addresses of their devices (and other personally identifiable information, \textit{e}.\textit{g}., email and a public key). During registration, users also specify their preferred modality through which the system can communicate with the user (\textit{e}.\textit{g}., email and/or push messages to the user device). Such communication is used during the notification phase.


%
%

%
%
%

\begin{figure*}[t]
\B
	\begin{center}
	\includegraphics[scale=0.45]{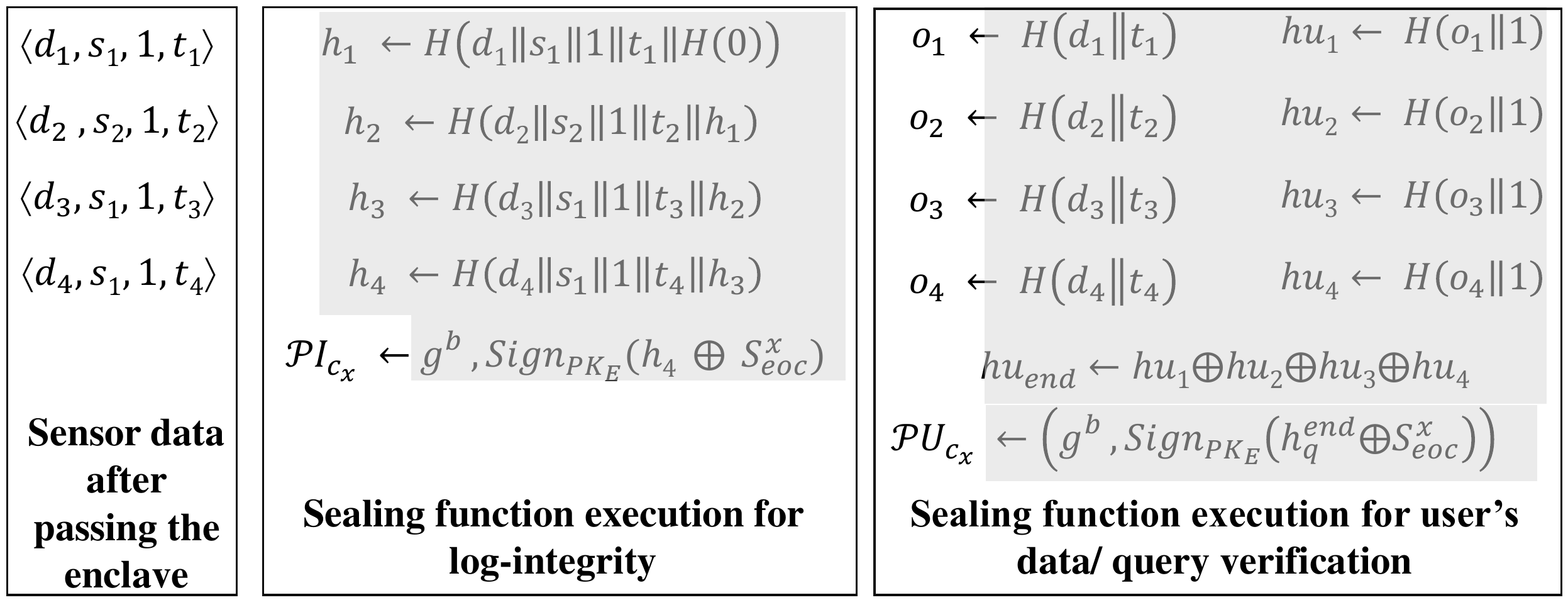}
	\end{center}
	\B
    \caption{Cryptographically sealing procedure executed on a chunk, $\mathcal{C}_x$. Gray-shaded data is not stored on the disk. White-shaded data is stored on the disk and accessible by SP. Figure shows proof-of-integrity for a chunk, $\mathcal{C}_x$.}
	\BBB
	\label{fig:sealing function execution}
\end{figure*}

\subsection{Notification Phase}
\label{subsec:notification phase}

The notification phase informs data-capture rules established by SP to the (registered) users by explicitly sending \emph{notice messages}. We consider two models for notification, differing based on acknowledgment from users.

In the \emph{notice-only model (NoM)}, SP informs users of data-capture rules, but users may not acknowledge receipt of the message. Such a model is used to implement policies, when data capture is mandatory, and the user cannot exercise control, over data capture. Since there is no acknowledgment, SP is only required to ensure that it sends a notice, but is not required to guarantee that the user received the notice. In contrast, a \emph{notice-and-ACK model (NaM)} is intended for discretionary data-capture rules that require explicit permission from users prior to data capture. Such rules may be associated, for instance, with fine-grained location services that require users' location. A user can choose not to let SP track his location, but will likely not be able to avail some services.

Implementation of notification differs based on the model used. Interestingly, since NaM requires acknowledgment, the notification phase is easier as compared to NoM that uses a trusted notifier to deliver the message to users. Below we discuss the implementation of both models:

\noindent\emph{\underline{Notification implementation in NoM}}. NoM assumes that, by default, data-capture rules are set not to retain any user data, unless SP, first, informs SGX about a data-capture rule, (\textit{i}.\textit{e}., SP cannot use the encrypted sensor data for building any application, see \encircle{9} in Figure~\ref{fig:Dataflow and computation in the protocol}). When SP creates a new data-capture rule, SP must inform SGX. Then, the enclave encrypts the data-capture rule using the public key (\textit{i}.\textit{e}., $\mathit{PK}_N$) of the notifier and informs the trusted notifier (via SP) about the encrypted data-capture rule by writing it outside of the enclave (in our user-case \S\ref{sec:introduction}, the university IT department works as a trusted notifier). Data-capture rules are maintained by SP on stable storage, which is read by SGX into the enclave to check, if the sensor data should be forwarded to SP. SGX can retain a cache of rules in the enclave, if such rules are still valid (and hence used for enforcement).\footnote{{\scriptsize Due to the enclave's limited memory, it cannot keep all valid and non-valid data-capture rules, after a certain size. Thus, the enclave writes all non-valid data-capture rules on the disk, after computing a secured hash digest over all rules. Taking a hash over the rules is needed to maintain the integrity of all rules.}} 
Finally, the trusted notifier acknowledges SP about receiving the encrypted data-capture rule, and then, informs users of the encrypted data-capture rule via signed notice messages. On receiving the notice message, the users may decrypt it and obtain the data-capture rule.

To see the role of \emph{trusted hardware} above, suppose that SP was responsible for informing users about data-capture rules directly. Since data-capture rules are also required by SGX during log-sealing (\textsc{Phase} 2), an adversarial SP may inform SGX, not to users, or may inform non-identical rules to users and to SGX. Hence, SP cannot inform the rule to users directly.

To see the role of  \emph{the trusted notifier} above, suppose that SP can directly inform users about encrypted data-capture rules obtained from SGX. An adversarial SP may not deliver the data-capture rule to all/some of the users; thus, an encrypted data-capture rule is not helpful. Thus, a trusted notifier ensures that the notice message is sent to all the registered users. Note that the trusted notifier might be a trusted web site that lists all the data-capture rules, which users can access.


\noindent\emph{\underline{Implementation of notification in NaM}}. Unlike NoM, the notification phase of NaM does not require the trusted notifier. In NaM, by default, SP cannot utilize all those sensor readings having device-ids for which the users have not acknowledged. Likewise NoM, in NaM, SP informs data-capture rules to SGX that encrypts the rule and writes outside of the enclave. The encrypted rules are delivered by SP to users, unlike NoM. On receiving the message, a user may securely acknowledge the enclave about her consent. The enclave retains all those device-ids that acknowledge the notice message for log-sealing phase and considers those device-ids during the log-sealing phase to retain their data while discarding data of others.



\subsection{Log Sealing Phase}
\label{sec:Log Sealing}

The second phase does cryptographically sealing of the sensor data for future verification against pre-notified data-capture rules. The sensor data is sealed into secured logs using authenticated data structures, \textit{e}.\textit{g}., hash-chains and XOR-linked lists (as shown in Figures~\ref{fig:sealing function execution},~\ref{fig:eoc}), by the sealing function, $\mathit{Sealing}(\mathit{PR}_E, \langle d_i, s_j, s_j.\mathit{state}, t_k\rangle)$, executed in the enclave at SP. Let us explain log-sealing in the context of WiFi connectivity data. The enclave reads the encrypted sensor data (\encircle{7} in Figure~\ref{fig:Dataflow and computation in the protocol}) and executes the three steps: (\textit{i}) decrypts the data, (\textit{ii}) checks the data against pre-notified valid data-capture rules, and (\textit{iii}) cryptographically seals the data and store \emph{appropriate secured logs}.

Below we explain our log sealing approach. To simplify the discussion, we consider the case when all the sensor data satisfies some data-capture rule (\textit{i}.\textit{e}., the state of all the sensor data is one), and hence, data is forwarded to and stored at SP~\S\ref{subsubsec:Sealing Entire Sensor Data}. Likewise, the protocol to deal with all sensor data having state one, a protocol can also deal with the case when some sensor data satisfies some data-capture rule, while remaining sensor data does not satisfy any rule (\textit{i}.\textit{e}., the state of the remaining sensor data is zero). However, due to page limitations, we skip details of such a protocol.


\medskip
\noindent\subsubsection{\textbf{{\underline{Sealing Entire Sensor Data}}}}
\label{subsubsec:Sealing Entire Sensor Data}
%
%
%

The sealing operation contains three phases: (\textit{i}) chunk creation, (\textit{ii}) hash-chain creation, and (\textit{iii}) proof-of-integrity creation; described below.

\medskip
\noindent\textbf{\textsc{Phase} 1: Chunk creation.} The first phase of the sealing operation finds an appropriate size of a chunk (to speed up the attestation process). Note that the incoming encrypted sensor data may be large, and it may create problems during verification, due to increased communication between SP and the verifier. Also, the verifier needs to verify the entire data, which have been collected over a large period of time (\textit{e}.\textit{g}., months/years). Further, creating cryptographic sealing over the entire sensor data may also degrade the performance of $\mathit{Sealing}()$ function, due to the limited size of SGX enclave. Thus, we first determine an appropriate chunk size, for each of which the sealing function is executed.

The chunk size depends on time epochs, the enclave size, the computational overhead of executing sealing on the chunk, and the communication overhead for providing the chunk to the verifier. A small chunk size reduces the communication overhead and maintains the log minimality property, thereby during the log verification phase, a verifier retrieves only the desired log chunks, instead of retrieving the entire sensor data. Consequently, SP stores many chunks.

\medskip
\noindent\textbf{\textsc{Phase} 2: Hash-chain creation.} Consider a chunk, $\mathcal{C}_x$, that can store at most $n$ sensor readings, each of them of the format: $\langle d_i, s_j, t_k\rangle$. The sealing function checks each sensor reading against data-capture rules and adds sensor state to each reading, as: $\langle d_i, s_j, s_j.\mathit{state}, t_k\rangle$. Since in this section we assumed that all sensor data will be stored, the sensor state of each sensor reading is set to 1. The sealing function starts with the first sensor reading of the chunk $\mathcal{C}_x$, as follows:

\noindent\textit{\underline{First sensor reading}}. For the first sensor reading of the chunk, the sealing function computes a hash function on value zero, \textit{i}.\textit{e}, $H(0)$. Then, the sealing function mixes $H(0)$ with the remaining values of the sensor reading, \textit{i}.\textit{e}., sensor-id, device-id, sensor state, and time, at which it computes the hash function, denoted by $H(d_1||s_j|| s_j.\mathit{state}|| t_k||H(0))$ that results in a hash digest, denoted by $h_1^x$. After processing the complete first sensor reading of the chunk $\mathcal{C}_x$, the enclave writes cleartext first sensor reading of $\mathcal{C}_x$, \textit{i}.\textit{e}., $\langle d_1,s_j, s_j.\mathit{state}, t_k\rangle$ on the disk, which can be accessed by SP.

\begin{figure*}[!t]
\BB
	\begin{center}
	\includegraphics[scale=0.45]{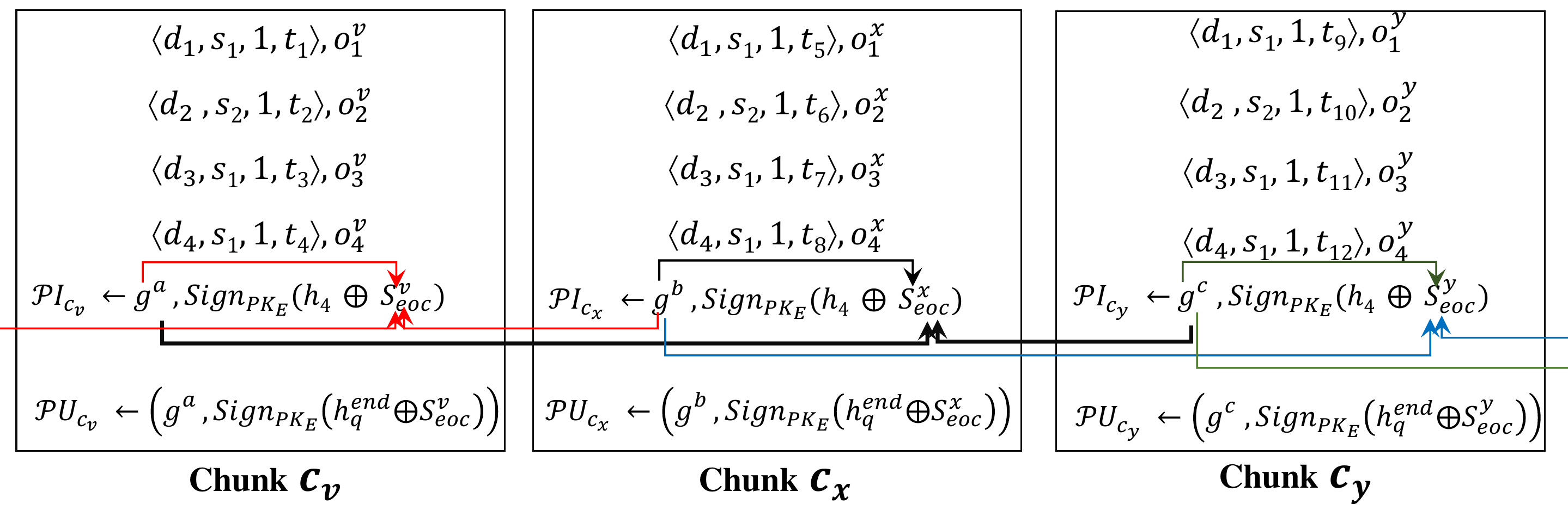}
	\end{center}
	\BB
    \caption{\textsc{Phase 3}: end of chunk, $S_{\mathit{eoc}}$, creation for three chunks. Observe that $S_{\mathit{eoc}}^{x} = g^a \oplus g^b \oplus g^c$.}
	\BBB
	\label{fig:eoc}
\end{figure*}

\noindent\textit{\underline{Second sensor reading}}. Let $\langle d_2, s_j, s_j.\mathit{state}, t_{k+1}\rangle$ be the second sensor reading. For this, the sealing function works identically to the processing of the first sensor reading. It computes a hash function on the second sensor values, while mixing it with the hash digest of the first sensor reading, \textit{i}.\textit{e}., $H(d_2||s_j|| s_j.\mathit{state}||t_{k+1}||h_1^x)$ that results in a hash digest, say $h_2^x$. Finally, the enclave writes the second sensor reading in cleartext on the disk.

\noindent\textit{\underline{Processing the remaining sensor readings}}. Likewise, the second sensor reading processing, the sealing function computes the hash function on all the remaining sensor readings of the chunk $\mathcal{C}_x$. After processing the last sensor reading of the chunk $\mathcal{C}_x$, the hash digest $h_n^x$ is obtained.

\medskip
\noindent\textbf{\textsc{Phase} 3: Proof-of-integrity creation.} Since each sensor reading is written on disk, SP can alter sensor readings, to make it impossible to verify log integrity by an auditor. Thus, to show that all the sensor readings are kept according to the pre-notified data-capture rules, the sealing function prepares an immutable proof-of-integrity for each chunk, as follows:

For each chunk $\mathcal{C}_i$, the sealing function generates a random string, denoted by $g^j$, where $i\neq j$. Let $\mathcal{C}_v$, $\mathcal{C}_x$, and $\mathcal{C}_y$ be three consecutive chunks (see Figure~\ref{fig:eoc}), based on consecutive sensor readings. Let $g^a$, $g^b$, and $g^c$ be random strings for chucks $\mathcal{C}_v$, $\mathcal{C}_x$, and $\mathcal{C}_y$, respectively. The use of random strings will ensure that any of the consecutive chunks have not been deleted by SP (will be clear in \S\ref{subsec:Attestation Phase}). Now, for producing the proof-of-integrity for the chunk $\mathcal{C}_x$, the sealing function: (\textit{i}) executes XOR operation on $g^a$, $g^b$, $g^c$, whose output is denoted by $S_{\mathit{eoc}}^{x}$, where $\mathit{eoc}$ denotes the end-of-chunk; (\textit{ii}) signs $h_n^x$ XORed with $S_{\mathit{eoc}}^{x}$ with the private key of the enclave; and (\textit{iii}) writes the proof-of-integrity for log verification of the chunk $\mathcal{C}_x$ with the random string $g^b$, as follows:
\centerline{\scriptsize
$\mathcal{PI}_{\mathcal{C}_x}=(g^b,\mathit{Sign}_{\mathit{PR}_E}(h_n^x \oplus S_{\mathit{eoc}}^x))$
}

\smallskip\noindent\textbf{Note.} We do not generate the proof for each sensor reading. The enclave writes only the proof and the random string for each chunk to the disk, which is accessible by SP. Further, the sensor readings having the state one are written on the disk, based on which SP develops services.

\smallskip\noindent\textbf{Example.} Please see Figure~\ref{fig:sealing function execution}, where the \textbf{middle box} shows \textsc{Phase} 2 execution on four sensor readings. Note that the hash digest of each reading is passed to the next sensor reading on which a hash function is computed with the sensor reading. After computing $h_4$, the proof-of-integrity, $\mathcal{PI}$, is created that includes signed $h_4\oplus S_{\mathit{eoc}}^x$ and a random string, $g^b$.

\smallskip\noindent\textit{Note. $g^{\ast}$ for the first chunk.} The initialization of log sealing function requires an initial seed value, say $g^{\ast}$, due to the absence of $0^{\mathit{th}}$ chunk. Thus, in order to initialize the secure binding for the first chunk, the seed value is used as a substitute random string.

\medskip
\noindent\subsubsection{\textbf{\underline{Sealing Data for User Data/Service Verification}}}
\label{subsubsec:Sealing Data for Query Execution}
While capturing \emph{user-associated data}, users may wish to verify their user-associated data against notified messages. Note that \emph{the protocol presented so far requires entire cleartext data to be sent to the verifier to attest log integrity} (it will be clear soon in \S\ref{subsec:Attestation Phase}). However, such cleartext data transmission is not possible in the case of user-associated data verification, since it may reveal other users' privacy. Thus, to allow verification of user-associated data (or service/query result\footnote{{\scriptsize The users, who access services developed by SP (as mentioned in \S\ref{sec:introduction}), may also wish to verify the query results, since SP may tamper with the data to show the wrong results.}} verification), we develop a new sealing method, consists of the three phases: (\textit{i}) chunk creation, (\textit{ii}) hash-generation, and (\textit{iii}) proof-of-integrity creation. Chunk creation phase of this new sealing method is identical to the above-mentioned chunk creation phase 1; see \S\ref{subsubsec:Sealing Entire Sensor Data}. Below, we only describe \textsc{Phase 2} and \textsc{Phase} 3.

\medskip
\noindent\textbf{\textsc{Phase} 2: Hash-generation.} Consider a chunk, $\mathcal{C}_x$, that can have at most $n$ sensor readings, each of them of the format: $\langle d_i, s_j, s_j.\mathit{state}, t_k\rangle$. 
Our objective is to hide users' device-id and its frequency-count (\textit{i}.\textit{e}., which device-id is prominent in the given chunk). Thus, on the $i^{\mathit{th}}$ sensor reading, the sealing function mixes $d_j$ with $t_k$, and then, computes a hash function over them, denoted by $H(d_j||t_k)$ that results in a digest value, say $o_i$. Note that hash on device-ids mixed with time results in two different digests for more than one occurrence of the same device-id. Note that $o_i$ helps the user to know his presence/absence in the data during attestation, but it will not prove that tampering has not happened with the data. Then, the sealing function mixes $o_i$ with the sensor state (to produce a proof of sensor state) of the $i^{\mathit{th}}$ sensor reading, and on which it computes the hash function, denoted by $H(o_i||s_j.\mathit{state})$ that results in a hash digest, denoted by $\mathit{hu}_i^x$. After processing the $i^{\mathit{th}}$ sensor reading of the chunk $\mathcal{C}_x$, the enclave writes $o_i$ on the disk. After processing all the $n$ sensor readings of the chunk $\mathcal{C}_x$, the sealing function computes XOR operation on all hash digests, $\mathit{hu}_i^x$, where $1\leq i\leq n$: $\mathit{hu}_1^x\oplus \mathit{hu}_2^x \oplus \ldots \oplus \mathit{hu}_n^x$, whose output is denoted by $\mathit{hu}_{\mathit{end}}^x$. (Reason of computing $\mathit{hu}_{\mathit{end}}^x$ will be clear in \S\ref{subsec:Attestation Phase}).

\medskip
\noindent\textbf{\textsc{Phase} 3: Proof-of-integrity creation for the user.} The sealing function prepares an immutable proof-of-integrity for users, denoted by $\mathcal{PU}$, for each chunk and writes on the disk. Likewise, proof-of-integrity for entire log verification, $\mathcal{PI}$ (\S\ref{subsubsec:Sealing Entire Sensor Data}), for each chunk, the sealing function obtains $S_{\mathit{eoc}}$; refer to \textsc{Phase} 3 in \S\ref{subsubsec:Sealing Entire Sensor Data}. Now, for producing $\mathcal{PU}$ for the chunk $\mathcal{C}_x$, the sealing function: (\textit{i}) signs $\mathit{hu}^x_{\mathit{end}}$ XORed with $S_{\mathit{eoc}}^x$ with the private key of the enclave, and (\textit{ii}) writes the signed output with the random string of the chunk, $g^b$, as $\mathcal{PU}_{\mathcal{C}_x}$.
\centerline{\scriptsize
$\mathcal{PU}_{\mathcal{C}_x}=(g^b,\mathit{Sign}_{\mathit{PR}_E}(\mathit{hu}^x_{\mathit{end}} \oplus S_{\mathit{eoc}}^x))$
}

\smallskip
\noindent\textbf{Note.} The enclave writes hash digests, $o_i$ for each sensor reading, the proof for user verification, and the random string for each chunk on the disk. Of course, the sensor readings having the state one are written on the disk.

\smallskip
\noindent\textbf{Example.} Please see Figure~\ref{fig:sealing function execution}, where the \textbf{last box} shows \textsc{Phase} 2 execution on four sensor readings to obtain the proof-of-integrity for the user, $\mathcal{PU}$.

\subsection{Attestation Phase}
\label{subsec:Attestation Phase}
The attestation phase contains two sub-phases: (\textit{i}) key establishment between the verifier and SP to retrieve logs, and (\textit{ii}) verification of the logs. Due to space restrictions, we skip the key establishment phase. Here, we briefly describe the verification process at the auditor and/or the user. 

\smallskip
\noindent\textbf{Verification process at the auditor.} Recall that the auditor can verify any part of the sensor data. Suppose the auditor wishes to verify a chunk $\mathcal{C}_x$; see Figure~\ref{fig:eoc}. Hence, entire sensor data (the data written in first box of Figure~\ref{fig:sealing function execution}) of the chunk $\mathcal{C}_x$, random strings $g^a$, $g^b$, and $g^c$ (corresponding to the previous and next chunks of $\mathcal{C}_x$; see Figure~\ref{fig:eoc}), and proof-of-integrity $\mathcal{PI}_{\mathcal{C}_x}$ are provided to the auditor. The auditor performs the same operation as in \textsc{Phase} 2 of \S\ref{subsubsec:Sealing Data for Query Execution}. Also, the auditor computes the end-of-chunk string $S^x_{\mathit{eoc}}=g^a\oplus g^b\oplus g^c$. At the end, the auditor matches the results of $h^x_n \oplus S^x_{\mathit{eoc}}$ against the decrypted value of received $\mathcal{PI}_{\mathcal{C}_x}$, and if both the values are identical, then it shows that the entire chunk is unchanged.

Note that since SP transfers sensor readings of the chunk $\mathcal{C}_x$, random strings ($g^a$, $g^b$, and $g^c$) and $\mathcal{PI}_{\mathcal{C}_x}$ to the user, SP can alter any transmitted data. However, SP cannot alter the signed $\mathit{Sign}_{\mathit{PR}_E}(h_n^x \oplus S_{\mathit{eoc}}^x)$, due to unavailability of the private key of the enclave, $\mathit{PR}_E$, which was generated and provided by the trusted authority to the enclave. Thus, by following the above-mentioned procedure on the sensor readings of $\mathcal{C}_x$, any inconsistency created by SP will be detected by the auditor.

\smallskip
\noindent\textbf{Verification process at the user.} If the user wishes to verify his data in a chunk, say $\mathcal{C}_x$, the user is provided all hash digests computed over device-id and time ($o_i$, see the last box in Figure~\ref{fig:sealing function execution}), time, sensor state, random strings $g^a$, $g^b$, and $g^c$ (see Figure~\ref{fig:eoc}), and the proof $\mathcal{PU}$ by SP. Since, the user knows her device-id, first, the user verifies her occurrences in the data by computing the hash function on her device-id mixed with provided time values and compares against received hash digests. This confirms the user's presence/absence in the data. Also, to verify that no hash-digest is modified/deleted by SP, the user computes the hash function on the sensor state mixed with the received $o_i$ ($1\leq i\leq n$, where $n$ in the number of sensor readings in $\mathcal{C}_x$) and computes $\mathit{hu}_{\mathit{end}}^x = h_1^x\oplus h_2^x \oplus \ldots \oplus h_n^x$. Finally, the user computes $\mathit{hu}_{\mathit{end}}^x \oplus S^x_{\mathit{eoc}}$ and compares against the decrypted value of $\mathcal{PU}$. The correctness of this method can be argued in a similar manner to the correctness of the verification at the auditor.

\section{Experimental Evaluation}
\label{sec:Experimental Evaluation}
This section presents our experimental results on live WiFi data. We execute \textsc{IoT Notary} on a 4-core 16GB RAM machine equipped with SGX at Microsoft Azure cloud.

\smallskip
\noindent\textbf{Setup.} In our setup, the IT department at UCI is the trusted infrastructure deployer. It also plays the role of the trusted notifier (notifying users over emailing lists). At UCI, 490 WiFi sensors, installed over 30 buildings, send data to a controller that forwards data to the cloud server, where \textsc{IoT Notary} is installed. The cloud keeps cryptographic log digests that are transmitted to the verifier, while sensor data, qualifies data-capture rules, is ingested into realtime applications supported by TIPPERS. We use SHA-256 as the hashing algorithm and 256-bit length random strings in \textsc{IoT Notary}. We allow users to verify the data collected over the last 30minutes (min). 


\begin{wrapfigure}{r}{6cm}
\BB
    \begin{minipage}{.99\linewidth}
	   \centering
	   \includegraphics[scale=0.44]{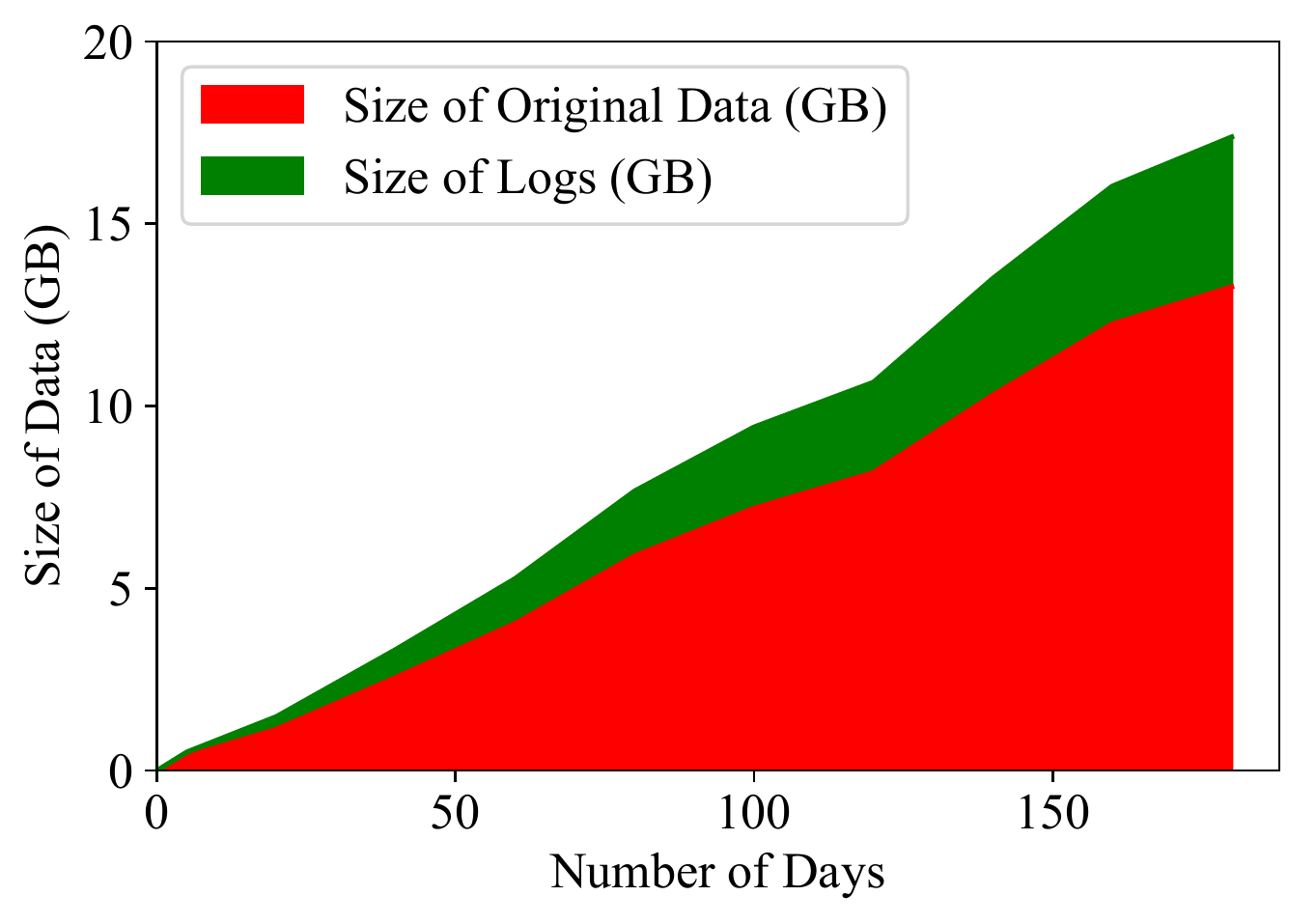}
				\caption{Exp 1: Storage overhead.}
				\label{fig:Storage overhead}
	\end{minipage}

    \begin{minipage}{.99\linewidth}
	   \centering
\includegraphics[scale=0.44]{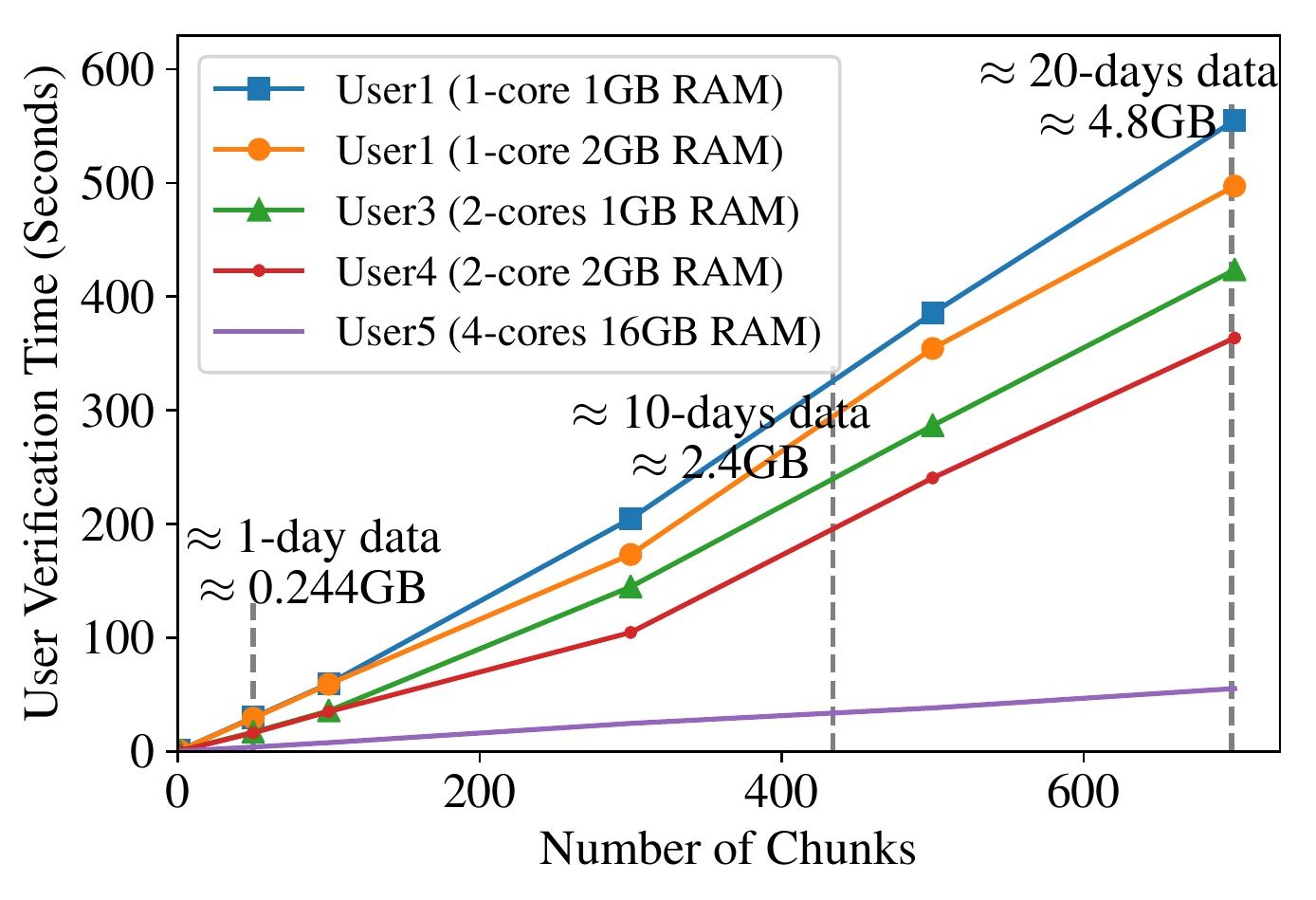}
				\caption{Exp 4: Verification time.}
				\label{fig:User verification time}
			\end{minipage}
\BBB
\end{wrapfigure}

\smallskip
\noindent\textbf{Dataset size.} Although \textsc{IoT Notary} deals with live WiFi data, we report results for data processed by the system over 180 days during which time \textsc{IoT Notary} processed 13GB of WiFi data having 110 million WiFi events.

\smallskip
\noindent\textbf{Data-capture rules.} We set the following four data-capture rules: (\textit{i}) \textit{Time-based}: always retain data, except from $t_i$ to $t_j$ each day; (\textit{ii}) \textit{User-location-based}: do not store data about specified devices if they are in a specific building; (\textit{iii}) \textit{User-time-based}: do not capture data having a specific device-id from $t_x$ to $t_y$ ($x\neq i$, $y\neq j$) each day; and (\textit{iv}) \emph{Time-location-based}: do not store any data from a specific building from time $t_x$ to $t_y$ each day. The validity of these rules was 40 days. After each 40-days, variables $i$, $j$, $x$, $y$ were changed.

\smallskip
\noindent\textbf{Exp 1. Storage overhead at the cloud.} We fix the size of each chunk to 5MB, and on average, each of them contains $\approx$ 37K sensor readings, covering around 30min data of 30 buildings in peak hours. Based on 5MB chunk size, we got 3291 chunks for 180 days. For each chunk, the sealing function generates two types of logs: (\textit{i}) for auditor verification that produced proof-of-integrity $\mathcal{PI}$ of size 512bytes, and (\textit{ii}) for user verification that produces hashed values (see Figure~\ref{fig:sealing function execution}) and proof-of-integrity for users $\mathcal{PU}$ of size 1.05MB. Figure~\ref{fig:Storage overhead} shows 180-days WiFi data size without having sealed logs (red color) and with sealed logs (green color).\footnote{{\scriptsize The reason of getting more chunks is that during non-peak hours 5MB chunk can store sensor readings for more than one hour. However, as per our assumption, we allow the user to verify the data collected over the last 30min. Hence, regardless of chunk is full or not, we compute the sealing function on each chunk after 30min.}}

\smallskip
\noindent\textbf{Exp 2. Performance at the cloud.} For each 5MB chunk, the sealing function took around 310ms to seal each chunk. This includes time to compute $\mathcal{PI}$, $\mathcal{PU}$ and encrypt them. 

%

\smallskip
\noindent\textbf{Exp 3. Auditor verification time.} The auditor at our campus has a 7th-Gen quad-core i7CPU and 16GB RAM machine. It downloads the chunks from the cloud and executes auditor verification. The auditor varied the number of chunks from 1 to 3000; see Table~\ref{tbl:auditor_verification_time_cloud}. Note that to attest one-day data across 30 buildings, the auditor needs to download at most 50 chunks, which took less than 1min to verify. Observe that as the number of chunks increases, the time also increases, due to executing the hash function on more data.

%
%
%

\smallskip
\noindent\textbf{Exp 4: Verification at a resource-constrained user.} To show the practicality of \textsc{IoT Notary} for resource-constrained users, we considered four types of users, differing on computational capabilities (\textit{e}.\textit{g}., available main memory (1GB/2GB) and the number of cores (1 or 2 cores)). Each user verified 1/10/20-days data; see Figure~\ref{fig:User verification time}. Note that verifying 1-day data, which is $\approx$ 50 blocks, at resource-constrained users took at most 30s. As the number of blocks increases, the computational time also increases, where the maximum computational time to verify 20-days data was $<$ 10min. As the days increase, so does data transmitted to the user, which spills over to disk causing an increased latency. Also, we execute the same experiment on a powerful user having 4-core and 16GB machine. Note that as the number of core and memory increase, it results in parallel processing and absence of disk data read. Thus, the computation time decreases (see user 5 in Figure~\ref{fig:User verification time}).

\begin{table}[!t]
    \centering
    \scriptsize
    \begin{tabular}{|l|l|l|l|l|l|l|l|}
    \hline
    Number of Chunks  & 1   &50 &  100 & 500 & 1000 &  3000\\ \hline
    $\approx$ duration (day) & 30-60min & 1-2 & 2-5 & 8-18 & 35-55 & 175\\ \hline
    Time (seconds) & 1 & 49& 102 & 544 & 1160 & 4400 \\ \hline
    \end{tabular}
    \caption{The auditor verification time. Duration varies due to different class schedules in buildings and working hours.}
    \label{tbl:auditor_verification_time_cloud}
    \BBB
    \end{table}

\section{Comparison with Existing Work}
\label{sec:Comparison with Existing Work}

We classify the related work in the area of IoT attestation into the following three categories:

\noindent\textbf{Attestation in the context of IoT.} The existing remote attestation protocols
verify the internal memory state of untrusted devices through a trusted remote verifier. For example, AID~\cite{atone} attests the internal state of neighboring devices through key exchange and proofs-of-non-absence. 
SEDA~\cite{atthree} attests embedded devices and provides the number of devices that pass attestation. 
Also, DARPA~\cite{attwo} and SANA~\cite{sana16} allow detection of physical attacks by using heartbeat messages and provide aggregate network attestation. 
In short, existing work cannot verify sensor data against the data-capture rules, except sensors' internal state. In contrast, \textsc{IoT Notary} does not deal with the verification of the internal state of sensors, since in our case, (WiFi access-point) sensors deployed by a trusted entity (\textit{e}.\textit{g}., the university IT department). Of course, cyberattacks are possible on sensors to maliciously record data and that can also be detected by \textsc{IoT Notary}.

\noindent\textbf{Attestation using secure hardware.}~\cite{midare} provided SGX-based attestation method for physical attacks on the sensor. 
Fiware~\cite{fiware} provides secure key management through key vault running in SGX. 
However,~\cite{midare,fiware} cannot verify any sensor data. Also, in~\cite{midare,fiware}, if data-capture rules are mis-notified to the user, SGX cannot detect any inconsistency. In contrast, \textsc{IoT Notary} does not deal with attacks on sensors, as well as, a specific key management protocol. However, \textsc{IoT Notary} can detect and discard the sensor data that does not comply with the notifications released earlier.


\noindent\textbf{Integrity verification.} 
~\cite{chef} proposed a privacy-preserving scheme based on zero-knowledge proofs to detect log-exclusion attacks. 
~\cite{bloompaper} proposed a Bloom tree that stores proof of logs at an untrusted cloud. 
vSQL~\cite{DBLP:conf/sp/ZhangGKPP17} may be used for verifying cleartext query results. However, these techniques cannot detect log deletion and incur significant overheads. For example, vSQL takes more than 4000 seconds to verify a SQL query. 
In contrast, \textsc{IoT Notary} provides complete security to sensor data and realtime data attestation approach. 

\section{Conclusion}
\label{sec:Conclusion}
This paper presented a framework, \textsc{IoT Notary} for sensor data attestation through cryptographically enforced log-sealing mechanisms to produce immutable proofs, used for log verification. We improve the na\"{\i}ve end-to-end encryption model, where retroactive verification is not provable. The user-data verification mechanism allows users to revoke services of the concerned IoT space. Thus, we empower the users with the right-to-audit instead of right-to-own the data captured by sensors. \textsc{IoT Notary} is a part of a real IoT system (TIPPERS) and provides verification on live WiFi data with almost no overheads on users.




\end{document}